\documentclass{elsarticle}

\usepackage{amssymb}
\usepackage{amsmath}
\usepackage{hyperref}
\usepackage{cleveref}
\usepackage{multirow}
\usepackage{graphicx}
\usepackage{subfigure}
\usepackage{tikz}
\usepackage{xspace}
\usepackage{relsize}
\usepackage{listings}
\usepackage{inconsolata}
\usepackage{color}

\newcommand{\code}[1]{\lstinline|#1|}

\protected\def\plusplus{{\nolinebreak[4]\hspace{-.05em}\raisebox{.4ex}{\relsize{-3}\bf ++}}\xspace}
\newcommand{\CXX}{{\rm C}\plusplus}
\newcommand{\www}[1]{\href{#1}{#1}}

\crefname{lstlisting}{listing}{listings}
\Crefname{lstlisting}{Listing}{Listings}

\definecolor{matplotlib1}{rgb}{0.12,0.47,0.71}
\definecolor{matplotlib2}{rgb}{1,0.5,0.05}

\journal{Journal of Parallel and Distributed Computing}

\begin{document}
\graphicspath{data}

\begin{frontmatter}

\title{
    Subdomain Deflation Combined with Local AMG:\\
    a Case Study Using AMGCL Library
}
\author{Denis Demidov\corref{c1},\fnref{f1}}
\cortext[c1]{Corresponding author}
\ead{dennis.demidov@gmail.com}
\address{
    Kazan Branch of Joint Supercomputer Center,
    Russian Academy of Sciences,
    Lobachevsky st. 2/31, 420111 Kazan, Russian Federation}
\address{
    Kazan Federal University,
    Kremlyovskaya st. 18, 420008 Kazan, Russian Federation}

\author{Riccardo Rossi,\fnref{f2}}
\address{
    Department of Civil and Environmental Engineering, Universitat
    Polit\`{e}cnica de Catalunya (UPC), Barcelona 08034, Spain}
\address{
    CIMNE --- Centre Internacional de M\`{e}todes Num\`{e}rics en Enginyeria,
    CIMNE, Barcelona 08034, Spain}

\fntext[f1]{Contribution of Dr. Demidov was
funded by the state assignment to the Joint supercomputer center of the
Russian academy of sciences for scientific research, project
No~0065-2018-00015.}
\fntext[f2]{Dr. Rossi acknowledges the financial support to CIMNE via the
CERCA Programme/Generalitat de Catalunya.}

\begin{abstract}
    The paper proposes a combination of the subdomain deflation method and
    local algebraic multigrid as a scalable distributed memory preconditioner
    that is able to solve large linear systems of equations. The implementation
    of the algorithm is made available for the community as part of an open
    source AMGCL library.  The solution targets both homogeneous (CPU-only) and
    heterogeneous (CPU/GPU) systems, employing hybrid MPI/OpenMP approach in
    the former and a combination of MPI, OpenMP, and CUDA in the latter cases.
    The use of OpenMP minimizes the number of MPI processes, thus reducing the
    communication overhead of the deflation method and improving both weak and
    strong scalability of the preconditioner. The examples of scalar,
    Poisson-like, systems as well as non-scalar problems, stemming out of the
    discretization of the Navier-Stokes equations, are considered in order to
    estimate performance of the implemented algorithm. A comparison
    with a traditional global AMG preconditioner based on a well-established
    Trilinos ML package is provided.
\end{abstract}


\end{frontmatter}

\section{Introduction}

The need to solve large and sparse linear systems of equations is ubiquitous
in engineering and physics.  Sparse linear solvers thus represent one of the
cornerstones of modern numerical methods, justifying the variety of alternative
techniques available for the purpose.  Direct solvers, based on some form of
matrix factorization are simply not viable beyond a certain size (typically of
the order of a few millions of unknowns~\cite{hogg2013new, henon2002pastix}),
due to their intrinsic memory requirements and sheer computational cost.
Iterative methods thus become the only feasible alternative in addressing such
large scale problems. These techniques typically fall in the class of
Krylov-subspace solvers~\cite{saad2003iterative} or of Algebraic Multigrid
algorithms (AMG)~\cite{brandt1985algebraic, ruge1987algebraic,
Stuben1999}.  Among the common problems
faced by Krylov approaches is their lack of robustness (compared to direct
alternatives) as well as their limited algorithmic scalability as the problem
size grows. Indeed a common feature of Krylov techniques is that the number of
iterations needed for convergence increases with the problem size, thus making
the methods less attractive for the solution of extremely large problems. On
the other hand, advantages of Krylov methods are their simplicity and
computational efficiency.  These methods also typically show very good strong
scalability for a given problem size (that is, the solution time decreases as
more processes are used).

Conversely, multigrid methods show excellent weak scalability, exhibiting a
convergence largely independent on the problem
scale~\cite{cleary2000robustness}. In turn, their limitation is related to
their higher complexity and to the need for an expensive hierarchy, which may
reduce the overall scalability.

In practice it is customary to blend the capabilities of multigrid and Krylov
techniques so as to preserve some of the advantages of
each~\cite{Trottenberg2001}. A number of successful  libraries exist
implementing different flavors of such blending for distributed memory
machines~\cite{yang2002boomeramg, ml-guide}.  Even though such implementations
were proven to be weakly scalable, their strong scalability remains
questionable.  This fact motivated the renowned interest in Domain
Decomposition (DD) methods~\cite{smith2004domain} as well as the rise of a new
class of approaches named \emph{deflation techniques}~\cite{vuik1999efficient}.
Deflation techniques are closely related to Balancing Domain Decomposition with
Constraints (BDDC) methods. However, it has  been shown that deflation
approaches are slightly advantageous~\cite{Nabben2006}.

Essentially all of the recent efforts in this area~\cite{FrankVuik2001} take on
the idea of constructing a multi level decomposition of the solution space,
understood as a very aggressive coarsening of the original problem. The use of
such coarser basis is proved sufficient to guarantee good weak scaling
properties, while implying a minimal additional computational complexity and
thus  good strong scaling properties.  An additional advantage of such
approaches is the ease of combining them \emph{in a modular way} with serial,
SMP, or GPGPU accelerated preconditioners, which allows to make use of the vast
number of existing local preconditioners in order to provide an effective
MPI-based solver.

The current paper focuses specifically on the combination of a deflation approach
with a local OpenMP-parallel AMG preconditioner.  We explore the use of a
constant and linearly-varying subspace defined on each of the MPI subdomains,
verifying their weak and strong scalability properties.

Although the methods employed are presented in some detail, the focus of the
contribution is on the software implementation within the open-source \CXX
library AMGCL\footnote{\www{https://github.com/ddemidov/amgcl}}.  AMGCL
provides several parallel backends, such as OpenMP or CUDA, which allows us to
consider advantages of hybrid MPI/OpenMP and MPI/CUDA parallelism.

The paper begins by briefly describing deflation-based techniques. Then we
introduce the cases of constant and linear deflation. Next, we detail the
characteristics of the local AMG preconditioner used on the subdomains.
Finally, in the benchmarks section, we prove the scalability of the developed
solver by performing both strong and weak scalability tests on both CPU-only
and GPU-accelerated systems. The well known Trilinos ML package~\cite{ml-guide}
is used in order to provide a baseline comparison with the traditional approach
based on a global AMG preconditioner.

\section{Subdomain deflation}

In order to provide a short description of the deflation approach, let us
consider a linear system of equations in the following form:
\begin{equation} \label{eq:original}
    \mathbf A \mathbf x = \mathbf b.
\end{equation}
Here the matrix $\mathbf A \in \mathbb{R}^{n \times n}$ is assumed to be
invertible.

The idea of deflation is to decompose the unknown $\mathbf x$ into a fine scale
solution $\mathbf y$ and a coarse scale solution $\mathbf{\bar x} = \mathbf Z
\mathbf \lambda$, so that
\begin{equation} \label{eq:decomposed}
    \mathbf x = \mathbf y + \mathbf Z \mathbf \lambda.
\end{equation}
Here $\mathbf Z \in \mathbb{R}^{n \times m}$ is a rectangular matrix whose
columns contain the basis vectors of a subspace of size $m$.

By premultiplying \eqref{eq:original} with $\mathbf Z^T$ and using
\eqref{eq:decomposed} we may write
\begin{equation} \label{eq:3}
    \mathbf Z^T \mathbf A \left( \mathbf y + \mathbf Z \mathbf \lambda \right)
    = \mathbf Z^T \mathbf b.
\end{equation}
Substituting \eqref{eq:decomposed} into \eqref{eq:original}, and using
\eqref{eq:3} the original problem may be expressed in matrix form as
\begin{equation} \label{eq:equivalent}
    \begin{pmatrix}
        \mathbf A & \mathbf{A}\mathbf Z \\
        \mathbf Z^T \mathbf A & \mathbf Z^T \mathbf A \mathbf Z
    \end{pmatrix}
    \begin{pmatrix}
        \mathbf y \\
        \mathbf \lambda
    \end{pmatrix}
    =
    \begin{pmatrix}
        \mathbf b \\
        \mathbf Z^T \mathbf b
    \end{pmatrix}
\end{equation}

At this point it is customary to define the deflation matrix $\mathbf E =
\mathbf Z^T \mathbf A \mathbf Z$.  The matrix is invertible if $\mathbf A$ is
invertible and the columns of $\mathbf Z$ are linearly
independent~\cite{FrankVuik2001}.  Under this assumption we can eliminate
$\mathbf \lambda$ from the equations and obtain a modified system for $\mathbf
y$ equivalent to the original one.  Namely, this corresponds to expressing
$\mathbf \lambda$ as
\begin{equation} \label{eq:lambda_def}
    \mathbf \lambda = -\mathbf E^{-1} \mathbf A \mathbf Z \mathbf y + \mathbf E^{-1} \mathbf Z^T \mathbf b.
\end{equation}
Substituting \eqref{eq:lambda_def} into the first row of \eqref{eq:equivalent}
yields the following (singular) system:
\begin{equation}
    \left( \mathbf I - \mathbf A \mathbf Z \mathbf E^{-1} \mathbf Z^T \right)
    \mathbf A \mathbf y = \left( \mathbf I - \mathbf A \mathbf Z \mathbf E^{-1}
    \mathbf Z^T \right) \mathbf b
\end{equation}
which may be understood as an operator $P$ applied to both sides of the
following equality:
\begin{equation} \label{eq:deflated_sys}
    P\left( \mathbf A \mathbf y \right) = P\left( \mathbf b \right).
\end{equation}
Using the definition of $\mathbf E$, it may be easily shown that for any
vector~$\mathbf v$
\begin{equation}
    P\left( P\left( \mathbf v \right) \right) = P\left( \mathbf v \right),
\end{equation}
thus proving that the operator $P$ is mathematically a projection.  The overall
point is that instead of solving the original problem for the complete
solution, we can now solve \eqref{eq:deflated_sys} for $\mathbf y$ and recover
the original solution from \eqref{eq:decomposed} and~\eqref{eq:lambda_def}.

Depending on the choice of the basis in $\mathbf Z$ the system in
\eqref{eq:deflated_sys} may become easier to solve than the original.  In order
to get some intuition for why this may be true, think of constructing $\mathbf
Z$ by using the subspace spanned by the eigenvectors of $\mathbf A$ with lowest
associated eigenvalue. Such technique is called ``eigenvalue deflation'' and
its properties are well known in the mathematical community. Physically such
eigenvectors describe global features of the solution. In this case $\mathbf
\lambda$ would describe the solution of the original problem when projected
onto such space, making the vector $\mathbf{\bar x}$ a global approximation to
the solution. The correction $\mathbf y$ would then be associated only to the
remaining eigenvectors which are relatively ``rapidly varying''.  Hence, the
effect of deflation would be to remove the modes with low eigenvectors from the
system, thus improving the condition number and in turn improving the
convergence of Krylov solvers.

We can make a number of remarks at this point:
\begin{itemize}
    \item The solution $\mathbf y$ resides in a space orthogonal to the space
        described by $\mathbf Z$. Since the size of the problem $P\left(\mathbf
        A\right)$ is the same as the original problem~\eqref{eq:original}, this
        implies that the problem~\eqref{eq:deflated_sys} is singular. Krylov
        methods allow a solution of such system since the right-hand side
        $\mathbf b$ is also subjected to the same projection.
    \item If columns of $\mathbf Z$ are eigenvectors of $\mathbf A$, a choice
        known as ``eigenvalue deflation'', then it is easy to show that such
        eigenvectors and the respective eigenvalues are removed from the
        projected system. In other words, $P\left(\mathbf A\right)$ is
        ``deflated'' of such eigenvectors.  In particular, if the eigenvectors
        corresponding to the smallest eigenvalues were chosen for construction
        of $\mathbf Z$, then the smallest eigenvalues in the system would be
        removed, implying that the spectral radius of the projected system
        improves.
    \item If a slowly varying space is used in constructing $\mathbf Z$ then
        the same argument holds \emph{heuristically}, meaning that the spectral
        radius is improved as long as the deflation space used for construction
        of $\mathbf Z$ allows approximating the  eigenvectors with lower
        associated eigenvalues.
    \item The implementation of the projector $P$ depends on the inverse of
        matrix $\mathbf E$. If this inverse is only approximated then the
        application of the projector \emph{can only be understood as a left
        preconditioner applied to the original system}.
    \item Even after the application of the projector, the system in
        \eqref{eq:deflated_sys} is still large and sparse. Deflation does not
        change the subspace in which $\mathbf y$ resides, hence a
        preconditioner is still needed to allow the effective solution in terms
        of $\mathbf y$.
\end{itemize}

\begin{figure}
    \begin{center}
        \subfigure[Constant deflation]{
            \begin{tikzpicture}[scale=0.6]
                \draw[-,darkgray,line width=2pt] (0,0) -- (5,0);
                \draw[-,lightgray,line width=2pt] (5,0) -- (10,0);
                \draw[-,dashed] (0,0) .. controls (5,3) and (9,3) .. (10,0);
                \draw[-,matplotlib1,line width=1pt] (0,1.3) -- (5,1.3);
                \draw[-,matplotlib1,line width=1pt] (5,1.6) -- (10,1.6);
                \draw (2.5,-0.1) node[anchor=north]{domain 1};
                \draw (7.5,-0.1) node[anchor=north]{domain 2};
            \end{tikzpicture}
            \label{fig:def:const}
        }\quad
        \subfigure[Linear deflation]{
            \begin{tikzpicture}[scale=0.6]
                \draw[-,darkgray,line width=2pt] (0,0) -- (5,0);
                \draw[-,lightgray,line width=2pt] (5,0) -- (10,0);
                \draw[-,dashed] (0,0) .. controls (5,3) and (9,3) .. (10,0);
                \draw[-,matplotlib1,line width=1pt] (0,0.1) -- (5,2.2);
                \draw[-,matplotlib1,line width=1pt] (5,2.8) -- (10,0.5);
                \draw (2.5,-0.1) node[anchor=north]{domain 1};
                \draw (7.5,-0.1) node[anchor=north]{domain 2};
            \end{tikzpicture}
            \label{fig:def:lin}
        }
    \end{center}
    \caption{True (dashed black line) and deflated (solid blue line) solution
    in case of constant and linear deflation.}
    \label{fig:deflation}
\end{figure}
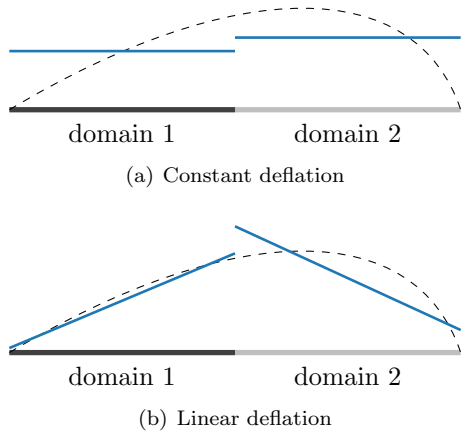

In practice the eigenvalues of $\mathbf A$ are generally not known.
Nevertheless it is possible to construct a global subspace by grouping the
unknowns which are ``close'' in the graph of $\mathbf A$, in a way similar to
what is done in algebraic multigrid.
Such choice is valid when the solution is geometrically smooth, which we assume
to be the case for all of the benchmark examples considered in the current
work. When such smoothness cannot be assumed, as happens for example in the
case of multiple materials or bubbly flows, the deflation idea can still be
exploited effectively, but some problem-specific knowledge is needed to group
the unknowns based on common physical characteristics~\cite{tang2006new}.
This is possible through the AMGCL user interface, however we do not target
this case in our benchmarks.  In practice, for smooth solutions one may
construct a basis by
assigning a value of 1 to one of these groups at a time while zeroing out other
groups, so that each deflation vector acts as an indicator for its respective
subdomain. This approach is known as ``subdomain
deflation''~\cite{FrankVuik2001}. \Cref{fig:def:const} provides a visual
demonstration of the approach: the ideal solution is approximated by a constant
solution over each MPI domain (two in the case shown) which represents the
effect of the term $\mathbf Z\mathbf \lambda$. Clearly the piecewise constant
solution only represents a rough approximation of the real solution. The term
$\mathbf y$ is thus interpreted as the correction needed to complete the
deflated solution and to arrive to the real one. The whole point here is that
the correction $\mathbf y$ is smaller than the total solution $\mathbf x$, it
is thus intuitive to understand how finding $\mathbf y$ requires a lesser
effort (read, less iterations) than finding the whole solution.
\Cref{fig:def:lin} shows how the idea can be further improved by allowing the
deflated solution to vary linearly within the domain. This choice provides
better approximation for the real solution, hence reducing the  residual and
requiring less iterations to convergence.
The drawback of this approach is the need to store the additional deflation
vectors and the increased size of the coarse system, the latter being more
important with the increasing number of MPI processes.

In the context of distributed linear algebra, it is customary that matrix
$\mathbf A$ is already mapped to MPI domains so as to preserve the locality of
entries in the graph. Basing on this observation one can assign one deflation
subdomain to each MPI process.

\subsection{On precision of the coarse system solution} \label{sec:inexactE}

The approach just shown consists in identifying a modified ``deflated'' system,
written in terms of an auxiliary variable $\mathbf y$ which is easier to solve
by iterative techniques. The core idea of the approach is that the real
solution $\mathbf x$ is easily recovered from $\mathbf y$ by performing a
simple correction step.  As observed in the remarks, such approach relies
heavily on the correct definition of the deflated system, that is, on the exact
application of the deflation projector.  In practice this implies that if we
were to compute an inexact approximation $\mathbf {\hat E} \approx \mathbf
E^{-1}$, then we would be solving the following system:
\begin{equation} \label{eq:approx}
    \left( \mathbf I - \mathbf A \mathbf Z \mathbf{\hat E}\mathbf  Z^T \right)
    \mathbf A \mathbf {\hat y} = \left( \mathbf I - \mathbf A \mathbf Z
    \mathbf{\hat E}\mathbf  Z^T \right) \mathbf b
\end{equation}
thus yielding a solution $\mathbf{\hat y}$ which only approximately complies
with the requirements. The immediate impact of this is that even after the
final correction term, the solution will depend on the local approximation
$\mathbf{\hat E}$ in a way that is difficult to quantify.  This is not
acceptable in a black box solver, since it does not guarantee the final level
of precision.  Hence, the application of the inverse of $\mathbf E$ must be
computed very accurately if the projector is intended to be precise. Such
requirement practically rules out the choice of Krylov or multigrid methods in
the solution of the coarse problem, since the cost of enforcing a very high
precision typically exceeds that of a direct approach for the problem sizes
that are typical in the coarse space.

Even though the current paper focuses mostly on the application of the exact
``projector''  the possibility exists to consider the deflation as
``inexact''~\cite{Tang2009}
thus allowing the use of a more relaxed tolerance in the solution of the coarse
problem. This implies considering the projector as a left preconditioner for
the original linear problem, to be combined multiplicatively with other
preconditioners. Such preconditioner should be considered as ``variable''
during the iterations (due to the inexact inverse of $\mathbf E$) so its
application might require the use of a flexible Krylov solver, for example the
flexible GMRES variant~\cite{saad2003iterative}.

\subsection{Domain decomposition approaches}

The practical importance of deflation  is in the context of domain
decomposition (DD) methods. The idea is that, in the context of distributed
linear algebra, the problem to be solved is divided into subdomains normally
obtained by grouping the unknowns so that they represent physically contiguous
portions of the domain to be solved.  Under such assumption, an appealing
possibility is to construct the deflation space in $\mathbf Z$ by assuming that
each of the basis vectors is purely local to one of the
subdomains~\cite{FrankVuik2001}. At this point one may construct a
preconditioner following a classic domain decomposition approach, that is,
discarding the columns of $\mathbf A$ corresponding to nonlocal entries. If the
deflation space is properly defined (for example taking it as constant over
each subdomain) then the solution $\mathbf y$ is discontinuous across the
subdomain borders, but such discontinuity may easily be fixed with local
corrections needed to retrofit the coarse solution to the real one. Such
corrections are of local nature, and tend to show ``high frequency'' behavior.
As such they are quickly addressed by local preconditioners.  It should be
intuitive how an increase in the number of subdomains improves the quality of
the deflation space and thus reduces the correction needed in computing
$\mathbf y$. This reflects in a reduction in the number of iterations needed to
have the overall problem to converge, thus combining optimally with local
preconditioners.

\subsection{Constant deflation}

The easiest option in constructing the deflation is named ``constant'' or
``subdomain deflation''~\cite{FrankVuik2001}. In this case the solution on each
of the subdomains is approximated with a constant.  If the size of the
system is $n$ and the number of subdomains considered is $m$, then $\mathbf Z
\in \mathbb{R}^{n \times m}$ can be constructed as
\begin{equation}
    \mathbf Z_{ij} = \begin{cases}
        1,\quad i \in S_j; \\ 0, \quad \text{otherwise};
    \end{cases}
\end{equation}
where $S_j$ is the set of unknowns belonging to the $j$-th subdomain.

Note that $\mathbf Z$ defined this way is completely local to each of the
subdomains. This means that in practice it is more convenient to define and
store just the local subblock $\mathbf Z^l$ of the deflation space. In case of
the constant deflation $\mathbf Z^l$ consist of a single column filled with
ones.

\subsection{Linear Deflation}

\newcommand{\xxi}{\phi_x^i}
\newcommand{\yyi}{\phi_y^i}
\newcommand{\zzi}{\phi_z^i}

In some cases it is possible to associate physical coordinates with each
unknown $i$ in the system. In a three dimensional space we could define the
coordinates as $\left(\xxi,\yyi,\zzi\right)$.  Then we could approximate the
solution on each of the subdomains with a plane and interpret $\mathbf \lambda$
as the plane equation coefficients. This means we can define the local subblock
$\mathbf Z^l$ as a four-column matrix with
\begin{equation}
    \mathbf Z_{ij}^l = \begin{cases}
        1, & j=1; \\
        \xxi - c_x, & j=2; \\
        \yyi - c_y, & j=3; \\
        \zzi - c_z, & j=4.
    \end{cases}
\end{equation}
Here $(c_x, c_y, c_z)$ are coordinates of the barycenter of the current
subdomain.

\section{Implementation}

The deflation approach we discussed in the previous sections was implemented
with AMGCL~--- an open source \CXX library for solution of large sparse linear
systems with algebraic multigrid (AMG) method~\cite{AMGCLRef, Trottenberg2001}.
The library is developed in the Supercomputer center of the Russian academy of
sciences and is published under liberal MIT license.

\begin{figure}
    \begin{center}
        \includegraphics[width=\textwidth]{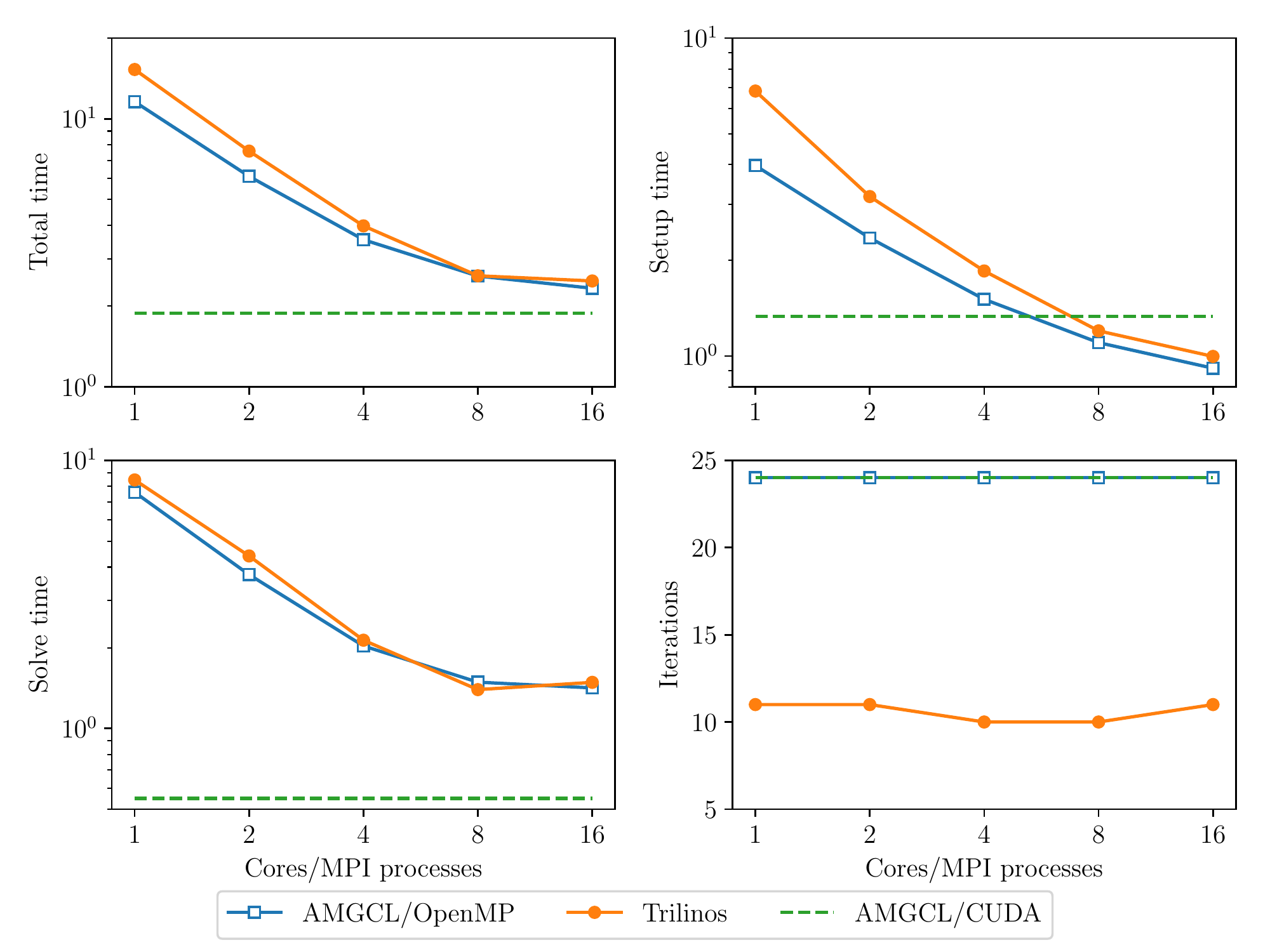}
    \end{center}
    \caption{Performance of OpenMP and CUDA backends of AMGCL vs Trilinos ML
    withing single compute node for solution of a 3D Poisson problem.}
    \label{fig:smem}
\end{figure}

AMGCL has minimal set of dependencies, and is designed in a generic and
extensible way that allows one to easily construct an AMG variant from
available algorithmic components of the method, such as coarsening strategy or
relaxation technique. The created hierarchy is transferred to one of the
provided backends for the solution phase.  Available backends support various
parallelization technologies such as OpenMP (for use with multicore CPUs), CUDA
or OpenCL (for use with modern GPUs). To illustrate this point, \cref{fig:smem}
shows OpenMP scalability of AMGCL for solution of a 3D Poisson problem on a
single cluster node. The problem is solved on a uniform $150^3$ mesh, the
system matrix contains 3~375~000 unknowns and 23~490~000 nonzero values. The
OpenMP-based AMGCL solution is compared with the MPI-based Trilinos ML solution,
where all MPI processes are allocated within the same compute node, and
with CUDA-based AMGCL solution, using NVIDIA Tesla K80 GPU available on the
same node. The node is a dual socket system with two Intel Xeon E5-2640 v3
CPUs. The OpenMP-based AMGCL solution shows performance comparable with Trilinos
ML. The AMGCL version uses one sweep of SPAI-0~\cite{broker2002sparse}
relaxation as smoother, and the Trilinos version uses two sweeps of
Gauss-Seidel relaxation. This explains the fact that AMGCL needs about twice
more iterations than Trilinos ML to converge, but the overall solution time is
comparable for both versions. The CUDA-based solution is able to outperform
both OpenMP-based AMGCL and Trilinos ML by about 2.5 times during solution
phase, and is overall faster by about 30\%.

The users of the library may easily
implement their own backends in order for the library to work transparently
with their custom data structures and algorithms. The main focus of AMGCL, as
opposed to many available non-commercial packages, is to provide a non-MPI,
``single-node'' implementation of the AMG. The proposed approach allows to
extend the library functionality onto MPI clusters in a modular way.

\begin{lstlisting}[float=t,
    caption={Solving linear system with AMGCL.}, label=lst:hello]
// Assemble system matrix in CRS format:
std::vector<int> ptr, col;
std::vector<double> val;
int n = assemble(ptr, col, val);

// Select the backend to use:
typedef amgcl::backend::builtin<double> Backend;     $\label{line:backend}$

// Define solver components:
typedef amgcl::make_solver<                          $\label{line:solver:def:begin}$
        amgcl::amg<                     // preconditioner:
            Backend,
            amgcl::coarsening::smoothed_aggregation, $\label{line:coarsening}$
            amgcl::relaxation::damped_jacobi         $\label{line:relaxation}$
            >,
        amgcl::solver::cg<Backend>      // iterative solver:
    > Solver;                                        $\label{line:solver:def:end}$

// Set solver parameters:
Solver::params prm;                                  $\label{line:prm:begin}$
prm.solver.tol = 1e-6;
prm.precond.relax.damping = 0.8;                     $\label{line:prm:end}$

// Setup solver:
Solver solve(boost::tie(n, ptr, col, val), prm);     $\label{line:setup}$

// The RHS and the solution vectors:
std::vector<double> rhs(n, 1.0), x(n, 0.0);          $\label{line:vectors}$

// Solve the system for the given RHS and initial solution:
int    iters;
double error;
boost::tie(iters, error) = solve(rhs, x);            $\label{line:solve}$
\end{lstlisting}

\Cref{lst:hello} shows a basic example of using the library.
Line~\ref{line:backend} selects the backend to use. Here we use the builtin
backend with \code{double} as a value type.  Lines~\ref{line:solver:def:begin}
to \ref{line:solver:def:end} define the solver type that we will use. The
\code{make_solver} class binds together two concepts: a preconditioner (in this
case an \code{amg} class), and an iterative solver (\code{cg}).
Lines~\ref{line:prm:begin} to \ref{line:prm:end} show how to set solver
parameters. Here we set the desired relative residual norm and a damping
parameter for the damped Jacobi relaxation algorithm chosen in
line~\ref{line:relaxation}. Most of the parameters have reasonable default
values, so we only need to change what is necessary. An instance of the solver
is constructed in line~\ref{line:setup} for the assembled sparse matrix in the
Compressed Row Storage (CRS) format. The instance is
then used to solve the system for the given right-hand side in
line~\ref{line:solve}.

Now, if we decided to use another backend, for example, the NVIDIA CUDA one, we
would only need to change the definition of the \code{Backend} type in
line~\ref{line:backend} from \code{amgcl::backend::builtin<double>} to
\code{amgcl::backend::cuda<double>} and replace the type of the vectors
\code{rhs} and \code{x} with \code{thrust::device_vector<double>} in
line~\ref{line:vectors}. As a result, the constructed solver would be
transferred to the supported GPU, and the solution would be performed on the
GPU. In a similar way, to change any of the algorithmic components of the
solver, we just need to adjust its type definition in
lines~\ref{line:solver:def:begin} to \ref{line:solver:def:end}.

This simple example already exposes some of the library design choices. AMGCL
uses policy-based design~\cite{gamma1995design, alexandrescu2001modern} for the
classes like \code{make_solver} or \code{amg}, so that individual algorithmic
components of the classes are selected at compile time by the corresponding
template parameters. The most important concepts of the library are backends,
iterative solvers, and preconditioners. A backend is a unified set of data
types and operations on them packaged into a class. Iterative solvers and
preconditioners are implemented in terms of backend operations, so that
backends are easily switched by changing a template parameter. The algebraic
multigrid class \code{amgcl::amg} is an implementation of the preconditioner
concept and in turn depends on concepts of coarsening and relaxation schemes.
In \cref{lst:hello} we use smoothed aggregation method as coarsening and damped
Jacobi method as relaxation (lines \ref{line:coarsening} and
\ref{line:relaxation}). Each component in the library defines a parameter
structure \code{params} complete with reasonable default values. Whenever a
parent component depends on a child, the child's parameters are included into
the parent's ones. For example, parameters for \code{amgcl::make_solver} are
defined as shown in \cref{lst:params}.  This allows to have generic constructor
interface for all classes in AMGCL and to seamlessly integrate various concepts
together.

\begin{lstlisting}[float=t,
    caption={Nested defintion of solver parameters.}, label=lst:params]
template <class PrecondType, class SolverType>
class make_solver {
    ...
    struct params {
        typename PrecondType::params precond;
        typename SolverType::params  solver;
    };
    ...
};
\end{lstlisting}

The compile-time definition of the solver components allows the compiler to
apply efficient optimization techniques, but in practice the complete
specification of the algorithm may be problem-dependent and impossible to
provide in advance. Thus AMGCL provides a runtime interface that allows the
user to select the algorithm components with runtime parameters. The runtime
solver identical to the one used in \cref{lst:hello} is shown in
\cref{lst:runtime}. The only component that still has to be defined at compile
time is the selected backend. An instance of \code{boost::property_tree::ptree}
class is now used for input parameters, but the parameters have the same
generic structure as before. An attractive advantage of the Boost.PropertyTree
library~\cite{schaling2014boost} is that it supports import of the parameters
from a number of file formats, such as XML or JSON.

\begin{lstlisting}[float=t,
    caption={Runtime definition of the AMGCL solver.}, label=lst:runtime]
typedef amgcl::make_solver<
        amgcl::runtime::amg<Backend>,
        amgcl::runtime::iterative_solver<Backend>
    > Solver;

boost::property_tree::ptree prm;
prm.put("solver.type", "cg");
prm.put("solver.tol", 1e-6);
prm.put("precond.coarsening.type", "smoothed_aggregation");
prm.put("precond.relax.type", "damped_jacobi");
prm.put("precond.relax.damping", 0.8);

Solver solve(boost::tie(n, ptr, col, val), prm);
\end{lstlisting}

The subdomain deflation approach in AMGCL is implemented using the concepts
described above. The components it depends on are a backend, an iterative
solver, a local preconditioner, and a direct solver (used to solve the coarse
deflated system). All of these may be readily reused from the core library, and
we only need to define the global projection method in order to finish the
implementation of the algorithm.

An important feature of the deflation approach is that each of the local
problems is well posed. Indeed, ignoring non-local matrix coefficients is
largely equivalent to applying Dirichlet conditions to all of the non-local
unknowns.  This allows the use of local preconditioners within every
computational domain, and also constitutes an important advantage with respect
to classical DD techniques in which the local problems often lack sufficient
boundary conditions.

The global projector, or the deflation operator, is implemented in a matrix
free fashion.  That is,
\begin{equation} \label{eq:projector}
    \mathbf r^* := \left( \mathbf I - \mathbf A \mathbf Z \mathbf E^{-1}
    \mathbf Z^T \right) \mathbf r
\end{equation}
is computed as
\begin{equation}
    \mathbf t_1 = \mathbf Z^T \mathbf r, \quad \mathbf t_2 = \mathbf
    E^{-1}\mathbf t_1, \quad \mathbf r^* = \mathbf r - (\mathbf A\mathbf Z)
    \mathbf t_2
\end{equation}
where $\mathbf t_*$ are temporary vectors used during the computations. It
should be noted that in our implementation the $\mathbf A\mathbf Z$ matrix
product and matrix $\mathbf E$ are computed explicitly for the sake of
performance and to be able to apply state-of-the art solvers to the solution of
the coarse matrix $\mathbf E$. Avoiding the explicit computation of $\mathbf E$
would be feasible if one goes for inexact solution strategies as discussed in
\cref{sec:inexactE}. This might be mandatory if extreme scalability is to be
achieved.

The fact that the deflation basis stored in $\mathbf Z$ is by construction
purely local to each of the subdomains allows us to optimize the operations in
the above matrix-free scheme and  reduce the communication cost of the
algorithm. The same optimization can also be applied to the computation of
$\mathbf E$ thus improving the overall efficiency of the process.

To reiterate, the algorithm considered in the current paper is equivalent to
solution of the system \eqref{eq:deflated_sys} preconditioned with block AMG.
Each block is constructed locally on each MPI process, and corresponds to a
square diagonal subblock of the system matrix $A$. There is no overlap between
the blocks, and the only communication is happening during the computation of
the global projector~\eqref{eq:projector}.

\section{Benchmarks}

In this section we demonstrate performance and scalability of the proposed
approach on the example of a Poisson problem and a Navier-Stokes problem in a
three dimensional space.  To provide a reference, we compare performance of the
AMGCL library with that of the well-known Trilinos ML package~\cite{ml-guide}.
The source code of the benchmarks is available at a GitHub repository%
\footnote{\www{https://github.com/ddemidov/amgcl\_benchmarks}}.

The benchmarks were run on MareNostrum~4\footnote{Barcelona, Spain,
\www{https://www.bsc.es/marenostrum/}}, and PizDaint\footnote{Lugano,
Switzerland, \www{http://www.cscs.ch/computers/piz\_daint/}} clusters which we
gained access to via PRACE program. The MareNostrum~4 cluster has 3\,456
compute nodes, each equipped with two 24 core Intel Xeon Platinum 8160 CPUs,
and 96 GB of RAM. The peak performance of the cluster is 6.2 Petaflops.  The
PizDaint cluster has 5\,320 hybrid compute nodes, where each node has one 12
core Intel Xeon E5-2690 v3 CPU with 64 GB RAM and one NVIDIA Tesla P100 GPU
with 16 GB RAM. The peak performance of the PizDaint cluster is 25.3 Petaflops.

\subsection{3D Poisson problem}

The first set of our  experiment uses the classical 3D Poisson
problem~\cite{mathews1970mathematical}.  Namely, we look for the solution of
the problem
\begin{equation} \label{eq:poisson}
    -\Delta u = 1,
\end{equation}
in the unit cube $\Omega = [0,1]^3$ with homogeneous Dirichlet boundary
conditions.

The AMGCL implementation uses a BiCGStab(2)~\cite{sleijpen1993bicgstab}
iterative solver preconditioned with subdomain deflation, as it
showed the best behaviour in our tests.  Smoothed aggregation AMG is used as
the local preconditioner. The Trilinos implementation uses a CG
solver~\cite{barrett1994templates} preconditioned with smoothed aggregation AMG
with default ``SA'' settings, or domain decomposition method with
default ``DD-ML'' settings~\cite{ml-guide}.

\begin{figure}
    \begin{center}
        \includegraphics[width=\textwidth]{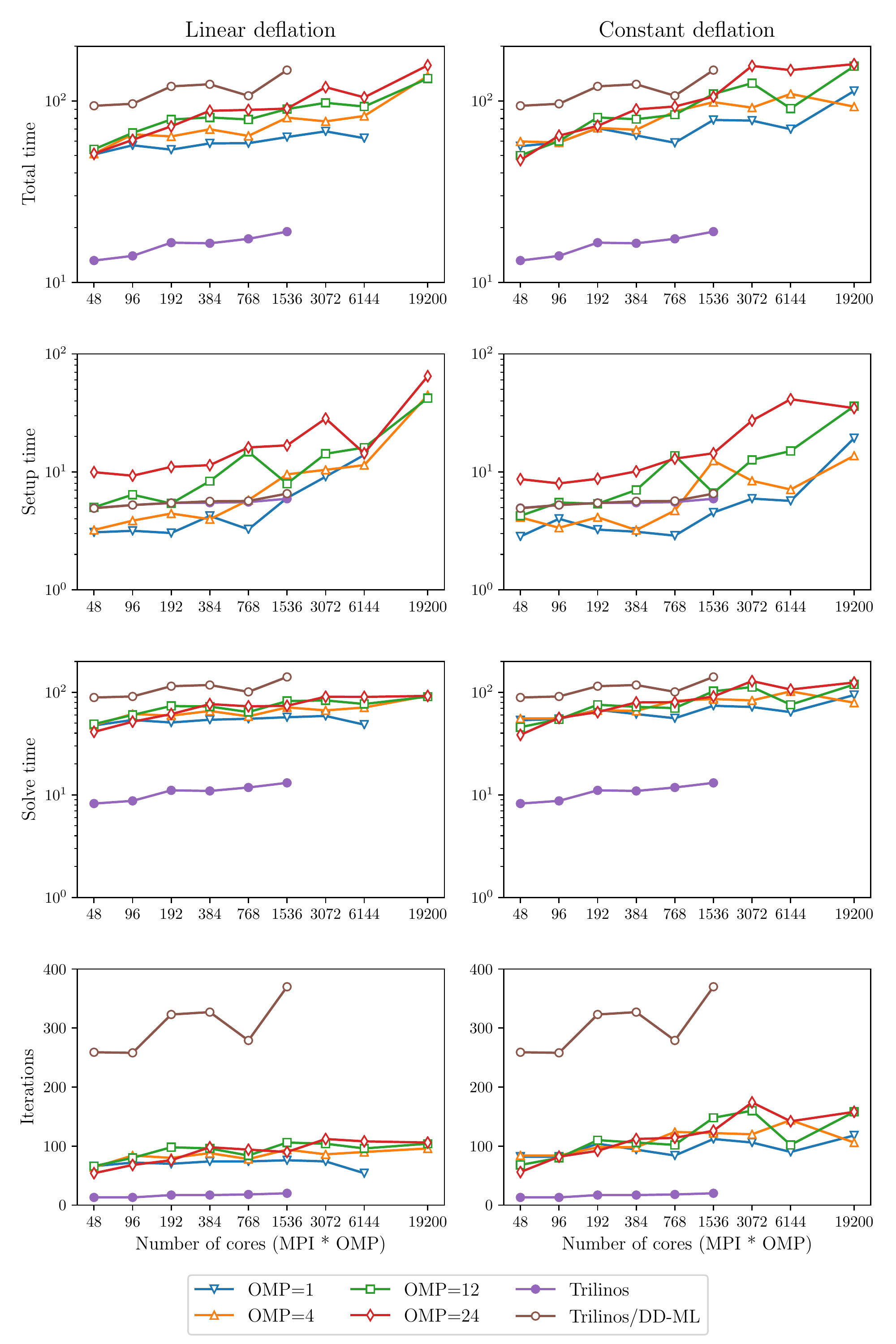}
    \end{center}
    \caption{Weak scalability of Poisson problem on the MareNostrum~4 cluster.}
    \label{fig:mn4:weak}
\end{figure}

\Cref{fig:mn4:weak} shows weak scaling of the solution on the MareNostrum~4
cluster. Here the problem size is chosen to be proportional to the number of
CPU cores with about~$100^3$ unknowns per core. The rows in the figure from top
to bottom show total computation time, time spent on constructing the
preconditioner, solution time, and the number of iterations.  The AMGCL library
results are labelled ``OMP=$n$'', where $n=1,4,12,24$ corresponds to the number
of OpenMP threads controlled by each MPI process. The Trilinos library uses
single-threaded MPI processes. The Trilinos data is only available for up to
1536 MPI processes, which is due to the fact that only 32-bit
version of the library was available on the cluster. The AMGCL data points for
19\,200 cores with ``OMP=1'' are missing because factorization of the deflated
matrix becomes too expensive for this configuration. AMGCL plots in the
left and the right columns correspond to the linear deflation and the constant
deflation correspondingly. The Trilinos and Trilinos/DD-ML lines
correspond to the smoothed AMG and domain decomposition variants accordingly
and are depicted both in the left and the right columns for convenience.

In the case of ideal scaling the timing plots on this figure would be strictly
horizontal. This is not the case here: instead, we see that both AMGCL and
Trilinos loose about 6-8\% efficiency whenever the number of cores doubles.
The proposed approach performs about three times worse that
the AMG-based Trilinos version, and about 2.5 times better than the domain
decomposition based Trilinos version. This is mostly governed by the number of
iterations each version needs to converge.

We observe that AMGCL scalability becomes worse at the higher number
of cores. We refer to the \cref{tab:prof:weak} for the explanation.  The table
presents the profiling data for the solution of the Poisson problem on the
MareNostrum~4 cluster. The first two columns show time spent on the setup of
the preconditioner and the solution of the problem; the third column shows the
number of iterations required for convergence. The ``Setup'' column is further
split into subcolumns detailing the total setup time and the time required for
factorization of the coarse system.  It is apparent from the table that
factorization of the coarse (deflated) matrix starts to dominate the setup
phase as the number of subdomains (or MPI processes) grows, since we use a
sparse direct solver for the coarse problem. This explains the fact that the
constant deflation scales better, since the deflation matrix is four times
smaller than for a corresponding linear deflation case.

The advantage of the linear deflation is that it results in a better
approximation of the problem on a coarse scale and hence needs less iterations
for convergence and performs slightly better within its scalability limits,
but the constant deflation eventually outperforms linear deflation as the scale
grows.

\begin{figure}
    \begin{center}
        \includegraphics[width=\textwidth]{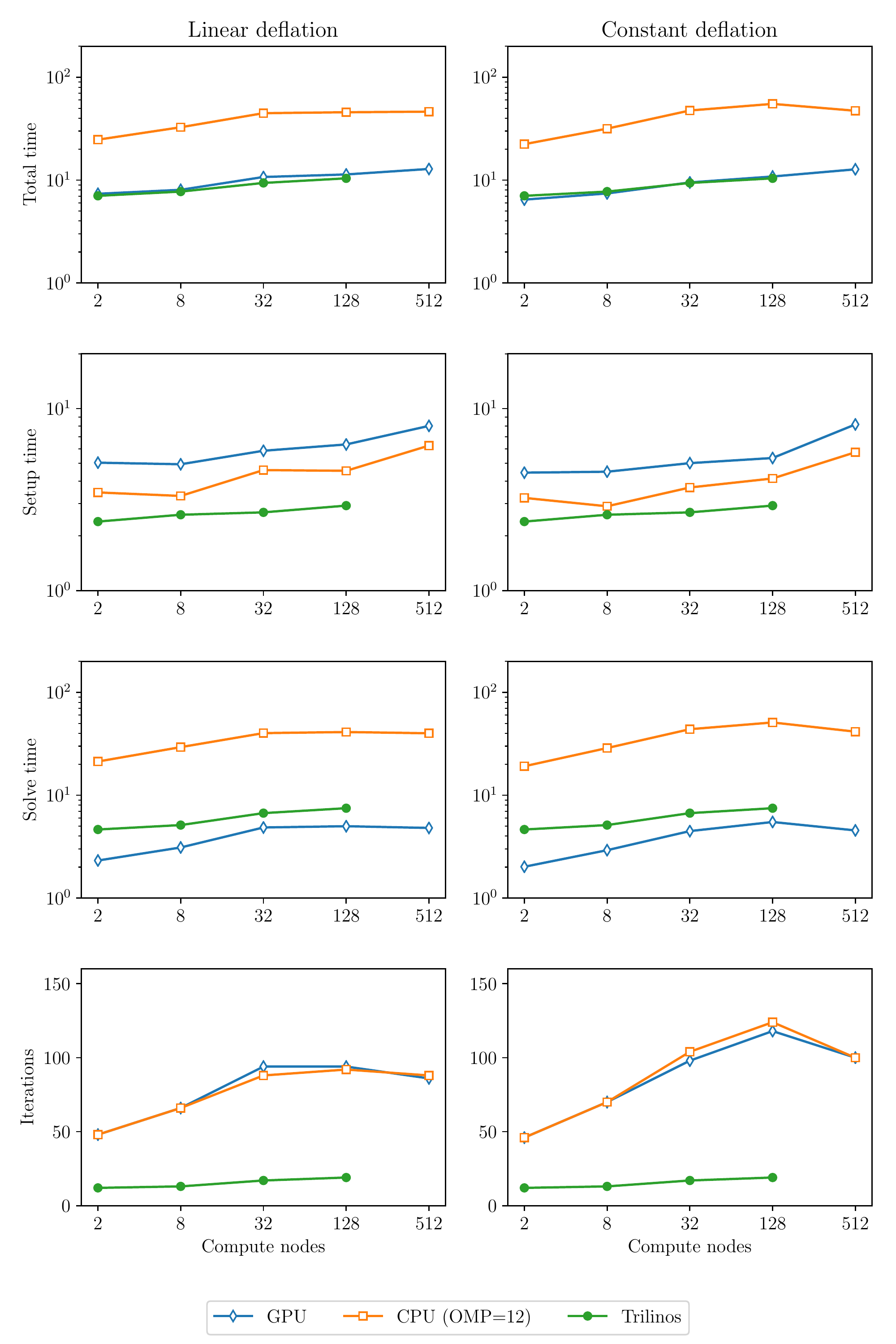}
    \end{center}
    \caption{Weak scalability of Poisson problem on the PizDaint cluster.}
    \label{fig:daint:weak}
\end{figure}

\Cref{fig:daint:weak} shows weak scaling of the Poisson problem on the PizDaint
cluster. The problem size here is chosen so that each node owns about $200^3$
unknowns. On this cluster we are able to compare performance of the
OpenMP and CUDA backends of the AMGCL library. Intel Xeon E5-2690 v3 CPU is
used with the OpenMP backend, and NVIDIA Tesla P100 GPU is used with the CUDA
backend on each compute node. The scaling behavior is similar to the
MareNostrum~4 cluster. We can see that the CUDA backend is about 9 times faster
than OpenMP during solution phase and 4 times faster overall. The discrepancy
is explained by the fact that the setup phase in AMGCL is always performed on
the CPU, and in the case of CUDA backend it has the additional overhead of
moving the generated hierarchy into the GPU memory. It should be noted that
this additional cost of setup on a GPU (and the cost of setup in general) often
can amortized by reusing the preconditioner for different right-hand sides.
This is often possible for non-linear or time dependent problems.  The
performance of the solution step of the AMGCL version with the CUDA backend
here is on par with the Trilinos ML package. Of course, this comparison is not
entirely fair to Trilinos, but it shows the advantages of using CUDA
technology.

\begin{table}
    \caption{Profiling data for the weak scaling of the Poisson problem on MareNostrum~4.}
    \label{tab:prof:weak}
    \begin{center}
    \begin{tabular}{|r|r|r|r|r|}
        \hline
        \multirow{2}{*}{Cores} & \multicolumn{2}{c|}{Setup} &
        \multirow{2}{*}{Solve} & \multirow{2}{*}{Iterations} \\ \cline{2-3}
        & Total & Factorize $E$ & & \\
        \hline
        \multicolumn{2}{|c}{OMP=1} & \multicolumn{3}{c|}{Linear deflation} \\
        \hline
        384     &  4.23 & 0.02 & 54.08 & 74 \\ % jobid=1413975
        1\,536  &  6.01 & 0.64 & 57.19 & 76 \\ % jobid=1413991
        6\,144  & 13.92 & 8.41 & 48.40 & 54 \\ % jobid=1414007
        \hline
        \multicolumn{2}{|c}{OMP=1} & \multicolumn{3}{c|}{Constant deflation} \\
        \hline
        384     & 3.11 & 0.00 & 61.41 &  94 \\ % jobid=1413976
        1\,536  & 4.52 & 0.01 & 73.98 & 112 \\ % jobid=1413992
        6\,144  & 5.67 & 0.16 & 64.13 &  90 \\ % jobid=1414008
        \hline
        \multicolumn{2}{|c}{OMP=12} & \multicolumn{3}{c|}{Linear deflation} \\
        \hline
        384     &  8.35 & 0.00 & 72.68 &  96 \\ % jobid=1413979
        1\,536  &  7.95 & 0.00 & 82.22 & 106 \\ % jobid=1413995
        6\,144  & 16.08 & 0.03 & 77.00 &  96 \\ % jobid=1414011
        19\,200 & 42.09 & 1.76 & 90.74 & 104 \\ % jobid=1414019
        \hline
        \multicolumn{2}{|c}{OMP=12} & \multicolumn{3}{c|}{Constant deflation} \\
        \hline
        384     &  7.02 & 0.00 &  72.25 & 106 \\ % jobid=1413980
        1\,536  &  6.64 & 0.00 & 102.53 & 148 \\ % jobid=1413996
        6\,144  & 15.02 & 0.00 &  75.82 & 102 \\ % jobid=1414012
        19\,200 & 36.08 & 0.03 & 119.25 & 158 \\ % jobid=1414020
        \hline
    \end{tabular}
    \end{center}
\end{table}

\begin{figure}
    \begin{center}
        \includegraphics[width=\textwidth]{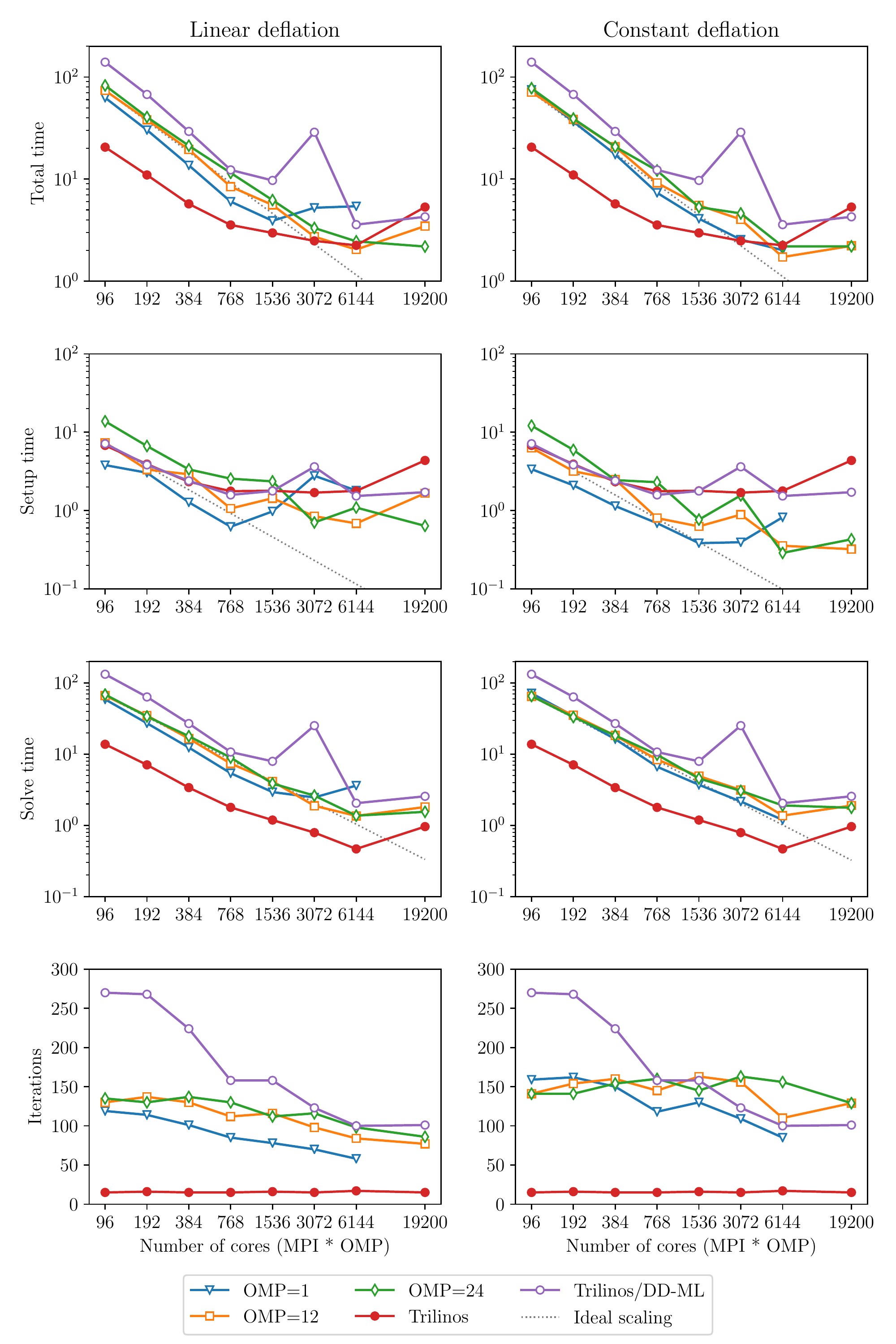}
    \end{center}
    \caption{Strong scalability of Poisson problem on the MareNostrum~4
    cluster.}
    \label{fig:mn4:strong}
\end{figure}

\Cref{fig:mn4:strong} shows strong scaling results for the MareNostrum~4
cluster. The problem size is fixed to $512^3$ unknowns
and ideally the compute time should decrease as we increase the number of CPU
cores.  The case of ideal scaling is depicted for reference on the plots with
thin gray dotted lines.

Here, AMGCL demonstrates scalability slightly better than that of the Trilinos
ML package. At 384 cores the AMGCL solution for OMP=1 is about 2.5 times slower
than Trilinos/AMG, and 2 times faster than Trilinos/DD-ML. As is expected for a
strong scalability benchmark, the drop in scalability at higher number of cores
for all versions of the tests is explained by the fact that work size per each
subdomain becomes too small to cover both setup and communication costs.

The profiling data for the strong scaling case is shown in
the \cref{tab:prof:strong}, and it is apparent that, as in the weak scaling
scenario, the deflated matrix factorization becomes the bottleneck for the
setup phase performance.

An interesting observation is that convergence of the method improves with
growing number of MPI processes. In other words, the number of iterations
required to reach the desired tolerance decreases with as the number of
subdomains grows, since the deflated system is able to describe the main
problem better and better.  This is especially apparent from the strong
scalability results (\cref{fig:mn4:strong}), where the problem size remains
fixed, but is also observable in the weak scaling case for ``OMP=1''
(\cref{fig:mn4:weak}).

\begin{table}
    \caption{Profiling data for the strong scaling of the Poisson problem on MareNostrum~4.}
    \label{tab:prof:strong}
    \begin{center}
    \begin{tabular}{|r|r|r|r|r|}
        \hline
        \multirow{2}{*}{Cores} & \multicolumn{2}{c|}{Setup} &
        \multirow{2}{*}{Solve} & \multirow{2}{*}{Iterations} \\ \cline{2-3}
        & Total & Factorize $E$ & & \\
        \hline
        \multicolumn{2}{|c}{OMP=1} & \multicolumn{3}{c|}{Linear deflation} \\
        \hline
        384     & 1.27 & 0.02 & 12.39 & 101 \\ % jobid=1404243
        1\,536  & 0.97 & 0.45 &  2.93 &  78 \\ % jobid=1404247
        6\,144  & 9.09 & 8.44 &  3.61 &  58 \\ % jobid=1404251
        \hline
        \multicolumn{2}{|c}{OMP=1} & \multicolumn{3}{c|}{Constant deflation} \\
        \hline
        384     & 1.14 & 0.00 & 16.30 & 150 \\ % jobid=1404244
        1\,536  & 0.38 & 0.01 &  3.71 & 130 \\ % jobid=1404248
        6\,144  & 0.82 & 0.16 &  1.19 & 85 \\ % jobid=1404252
        \hline
        \multicolumn{2}{|c}{OMP=12} & \multicolumn{3}{c|}{Linear deflation} \\
        \hline
        384     & 2.90 & 0.00 & 16.57 & 130 \\ % jobid=1404245
        1\,536  & 1.43 & 0.00 &  4.15 & 116 \\ % jobid=1404249
        6\,144  & 0.68 & 0.03 &  1.35 &  84 \\ % jobid=1404253
        19\,200 & 1.66 & 1.29 &  1.80 &  77 \\ % jobid=1404988
        \hline
        \multicolumn{2}{|c}{OMP=12} & \multicolumn{3}{c|}{Constant deflation} \\
        \hline
        384     & 2.49 & 0.00 & 18.25 & 160 \\ % jobid=1404246
        1\,536  & 0.62 & 0.00 &  4.91 & 163 \\ % jobid=1404250
        6\,144  & 0.35 & 0.00 &  1.37 & 110 \\ % jobid=1404254
        19\,200 & 0.32 & 0.02 &  1.89 & 129 \\ % jobid=1404989
        \hline
    \end{tabular}
    \end{center}
\end{table}

\begin{figure}
    \begin{center}
        \includegraphics[width=\textwidth]{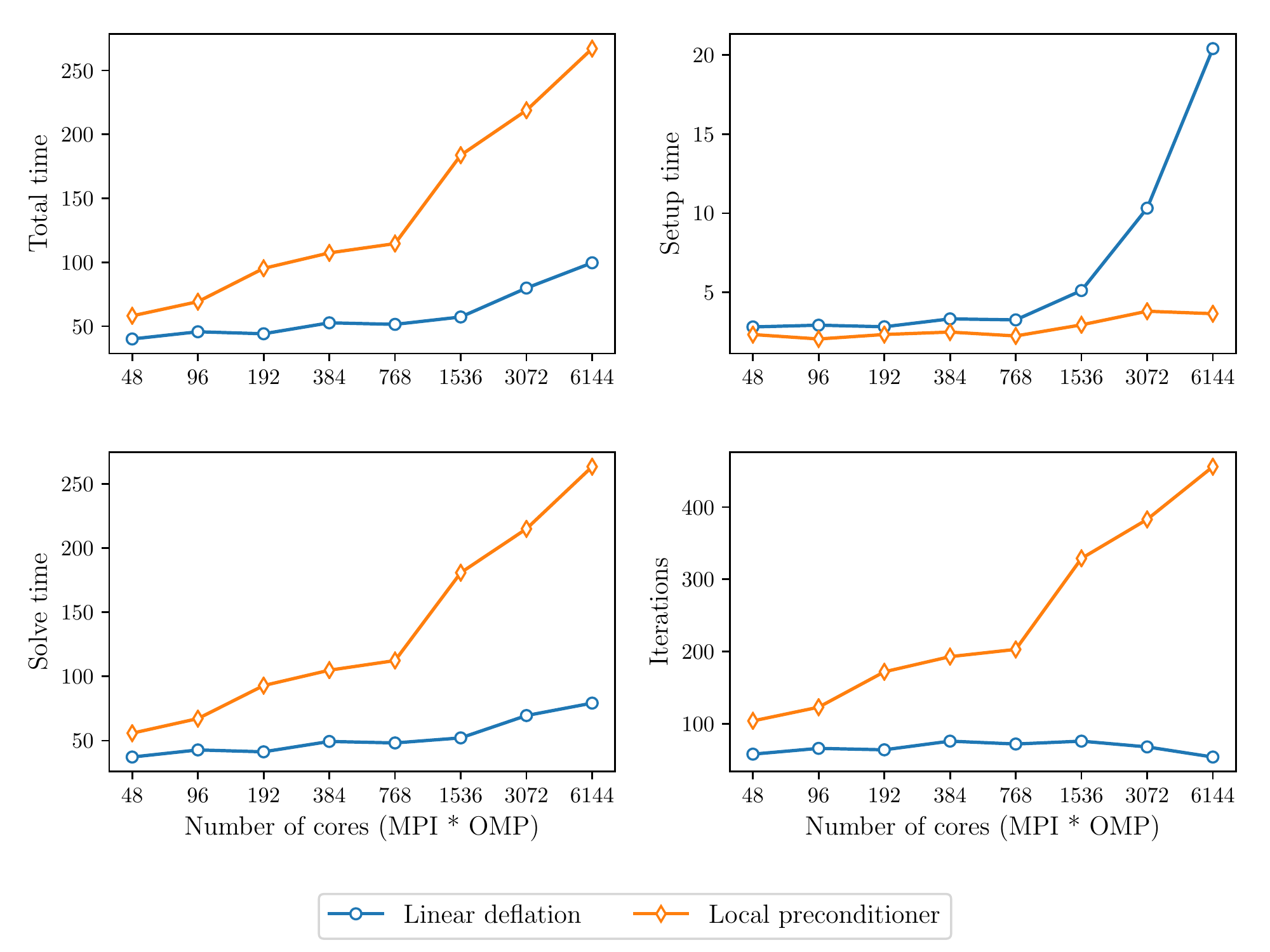}
    \end{center}
    \caption{Weak scaling of the Poisson problem. The proposed approach is
    compared with the case when AMG is used as a purely local preconditioner.
    OMP=1 in both cases.}
    \label{fig:mn4:deflation_vs_local}
\end{figure}

To conclude the benchmark, we compare the results of the proposed deflation
solver, which as we recall combines deflation and AMG as local preconditioner
(employing the AMGCL OpenMP version),
with the result that can be obtained by omitting the use of deflation and using
just the local AMG. As shown in \Cref{fig:mn4:deflation_vs_local} the use of
the local preconditioner alone is not weakly scalable, and leads to a number of
iterations that grows, approximately linearly with the problem size. On the
contrary, the use of deflation results in a flat number of iterations. Clearly
such improvement in the iteration count comes at the cost of solving a global
coarse problem, thus implying that each deflation iteration is slightly more
costly than applying the local AMG preconditioner. The timings however suggest
that the price is reasonably low, and that the use of deflation pays off.

\subsection{Navier-Stokes problem}

The second test problem is an incompressible Navier-Stokes problem discretized
on a non uniform 3D mesh with a finite element method:
\begin{subequations} \label{eq:nstokes}
\begin{gather}
    \frac{\partial \mathbf u}{\partial t} + \mathbf u \cdot \nabla \mathbf u  +
    \nabla p = \mathbf b, \\
    \nabla \cdot \mathbf u = 0.
\end{gather}
\end{subequations}
The problem is discretized using an equal-order tetrahedral Finite Elements,
stabilized employing an ASGS-type (algebraic subgrid-scale)
approach~\cite{Donea2003}. This results in a discretized linear system of
equations with a block structure of the type
\begin{equation} \label{eq:discrete_nstokes}
\begin{pmatrix}
    \mathbf K & \mathbf G \\
    \mathbf D & \mathbf S
\end{pmatrix}
\begin{pmatrix}
    \mathbf u \\ \mathbf p
\end{pmatrix}
=
\begin{pmatrix}
    \mathbf b_u \\ \mathbf b_p
\end{pmatrix}
\end{equation}
where each of the matrix subblocks is a large sparse matrix, and the blocks
$\mathbf G$ and $\mathbf D$ are non-square.  The overall system matrix for the
problem was assembled in the Kratos%
\footnote{\www{http://www.cimne.com/kratos/}} multi-physics
package~\cite{Dadvand2010,Dadvand2013} developed in CIMNE, Barcelona. The
matrix contains 4\,773\,588 unknowns and 281\,089\,456 nonzeros.
Such problem is routinely solved within Kratos by employing the ML solver with
the default ``NSSA'' settings. From the developer experience, this constitutes the
fastest ``out-of-the-box'' solver option, which is used here as the reference
line. Here we explored the use of the Schur complement solver capabilities
available within AMGCL.
The pressure variables are identified by a ``pressure-mask'' array, and the
AMGCL solver constructs an inexact pressure-Schur
preconditioner~\cite{Turek1999,Elman2005,Gmeiner2016}. The logical matrix
blocks in \eqref{eq:discrete_nstokes} are explicitly assembled for the sake of
performance. Each application of the preconditioner corresponds to the
following steps:
\begin{enumerate}
    \item Inexact solve of $\mathbf K \mathbf{ \hat u} = \mathbf b_u - \mathbf
        G \mathbf p$.
    \item Inexact matrix-free solve of $\left(\mathbf S - \mathbf D
        \mathop{\mathrm{diag}}\left(\mathbf K\right)^{-1}\mathbf G\right)\mathbf p = \mathbf b_p -
        \mathbf D \mathbf{\hat u}$.
    \item Inexact solve of $\mathbf K \mathbf{ u} = \mathbf b_u - \mathbf G
        \mathbf p$.
\end{enumerate}
A purely local  SPAI-0 (sparse approximate inverse)
preconditioner~\cite{broker2002sparse} is employed in the solution of the first
and third step.  Deflation in combination with local AMG is employed as a
preconditioner in the second step (the solution of the pressure Schur
complement). Here we choose $\mathbf M \approx \mathbf S^{-1}$ so that each
application of the preconditioner corresponds to an approximate solution of a
linear problem having $\mathbf{S}$ as a system matrix. Since a low tolerance is
targeted in the application of such preconditioner, a flexible GMRES (FGMRES)
solver is employed in the solution of the matrix-free problem, to ensure that a
preconditioner delivering a varying level of precision can be correctly
handled.  FGMRES is also used in the top level in targeting the overall
solution procedure.  The complete preconditioner is composed from the reusable
components provided by the AMGCL library, and \cref{lst:schurpc} shows the
definition of the resulting solver class as used in our benchmarks. The class
is using runtime interface, and \cref{lst:schur:json} shows JSON file
containing AMGCL options used for the tests.

\begin{lstlisting}[
    float=t,
    caption={AMGCL class definition for the Navier-Stoker solver.},
    label=lst:schurpc
]
typedef
    amgcl::mpi::make_solver<
        amgcl::mpi::schur_pressure_correction<          // top level preconditioner
            amgcl::mpi::make_solver<                    //   flow block solver
                amgcl::mpi::block_preconditioner<
                    amgcl::runtime::relaxation::as_preconditioner<Backend>
                    >,
                amgcl::runtime::iterative_solver
                >,
            amgcl::mpi::subdomain_deflation<            //   pressure block solver
                amgcl::runtime::amg<Backend>,
                amgcl::runtime::iterative_solver,
                amgcl::runtime::mpi::direct_solver<double>
                >
            >,
        amgcl::runtime::iterative_solver                // top level iterative solver
        >
    Solver;
\end{lstlisting}

\begin{figure}
    \begin{center}
        \includegraphics[width=\textwidth]{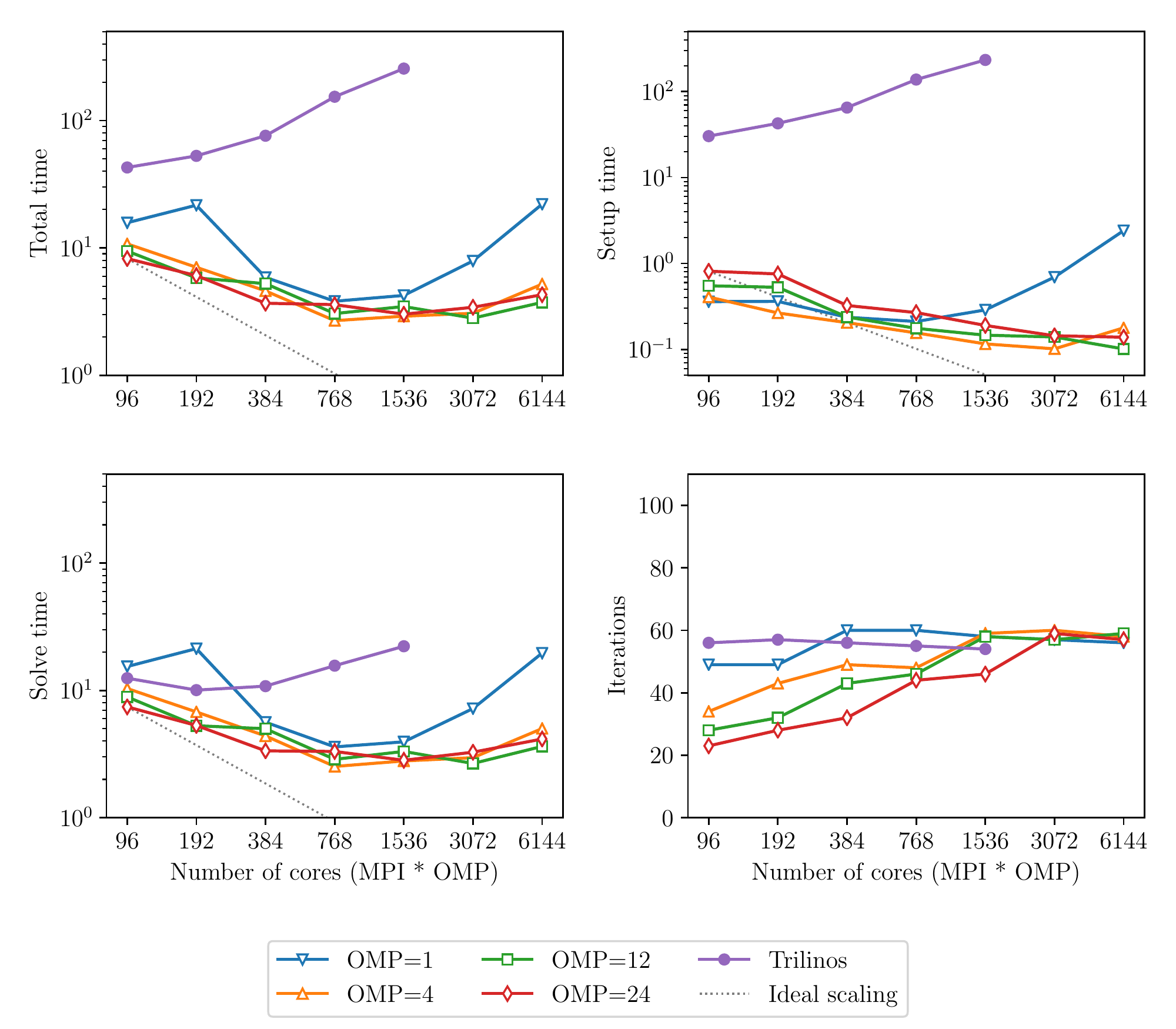}
    \end{center}
    \caption{Strong scalability of Navier-Stokes problem on the MareNostrum~4
    cluster.}
    \label{fig:mn4:schur}
\end{figure}

\Cref{fig:mn4:schur} shows scalability results for solving
problem~\eqref{eq:nstokes} on the MareNostrum~4 cluster. Since we are solving a
fixed-size problem, this is essentially a strong scalability test.  It should
be noted that Trilinos ML~\cite{ml-guide} does not provide field-split type
preconditioners, which is why the Trilinos benchmark was performed using the
non-symmetric smoothed aggregation variant (NSSA) applied to the monolithic
problem. Default NSSA parameters were employed in the tests.

Both AMGCL and ML preconditioners deliver a very flat number of iterations with
growing number of MPI processes. As expected, the pressure-Schur approach pays
off and the tested approach performs better than the monolithic approach in the
solution of the problem.  Overall the AMGCL implementation shows a decent,
although less than optimal parallel scalability. This is not unexpected since
the problem size quickly becomes too little to justify the use of more parallel
resources (note that at 192 processes, less than 25\,000 unknowns are assigned
to each MPI subdomain). Unsurprisingly, in this context the use of OpenMP
within each domain pays off and allows delivering a greater level of
scalability.

\begin{lstlisting}[
    float=t,
    caption={JSON file with AMGCL options used to solve the Navier-Stokes test
             problem.},
    label=lst:schur:json
]
{
    "solver": {
        "type" : "fgmres",
        "M" : 50,
        "tol" : 1e-4
    },
    "precond": {
        "usolver": {
            "solver": {
                "type" : "gmres",
                "tol" : 1e-3,
                "maxiter" : 5
            }
        },
        "psolver": {
            "isolver": {
                "type" : "fgmres",
                "tol" : 1e-2,
                "maxiter" : 20
            },
            "local" : {
                "coarse_enough" : 500
            }
        }
    }
}
\end{lstlisting}

\section{Conclusion}

The work presents a scalable preconditioner for sparse distributed linear
systems in form of a combination of the subdomain deflation method and local
algebraic multigrid. The implementation of the algorithm is included as a
reusable component into the open source AMGCL library. The results presented
prove that the solver enjoys a good weak scalability up to 19\,200 cores, the
maximum core count available during the testing. Strong scalability tests show
the solver efficiency for fixed size problems. The solution targets both
homogeneous (CPU-only) and heterogeneous (CPU/GPU) systems, employing hybrid
MPI/OpenMP approach in the former and MPI/CUDA in the latter cases. The use of
a heterogeneous cluster with GPUs installed on each node allows to gain an
overall 4x speedup for the complete solution of the benchmark problem under
consideration. The speedup factor results from a much faster solution phase
(around 10x speedup) and of a slightly slower setup phase.  The AMGCL solver,
both with OpenMP and CUDA backends, proves to be competitive with the Trilinos
ML package.

The use of OpenMP in combination with MPI proves important in achieving
scalability at high core counts. The reasons for such behaviour are easily
identified in the smaller size of the deflation matrix $\mathbf E$ to be
inverted, and in the minor MPI communication volume. Taking advantage of such a
feature, we believe that larger core counts are possible. For example, by using
48 cores per MPI on MareNostrum~4 one may target using 100K processor cores
with a deflation matrix of size 2000, easily within the scalable range.
However, such extreme scalability scenarios are still untested and will require
some additional effort in ironing out performance bottlenecks.  Targeting such
scenarios will be the focus of future works.

\section{Acknowledgments}

The authors thankfully acknowledge the support of the PRACE program (project
2010PA4058), in providing access to the MareNostrum~4 and PizDaint clusters.
Without such resources the testing would not have been possible.  The help of
Prof. Labarta of the POP Center of Excellence in improving the NUMA scalability
of the solver is also gratefully acknowledged.

\bibliographystyle{elsarticle-num}
\bibliography{ref}

\end{document}